\documentclass[iop]{emulateapj}
\usepackage{rotating}
\usepackage{apjfonts}
\usepackage{graphicx}
\usepackage{multirow}
\usepackage{comment}
\usepackage{natbib}
\usepackage{hyperref}
\usepackage{amsmath}
\usepackage{fontenc}
\usepackage{color}
\usepackage{indentfirst}
\begin{document}

\title{NEW INSIGHTS INTO THE FORMATION OF THE BLUE MAIN SEQUENCE IN
  NGC 1850}

\shorttitle{NGC 1850} 
\shortauthors{Y. Yang et al.}
\author{Yujiao Yang\altaffilmark{1,2}, Chengyuan Li\altaffilmark{3},
  Licai Deng\altaffilmark{2}, Richard de Grijs\altaffilmark{3,1,4},
  and Antonino P. Milone\altaffilmark{5,6}}

\altaffiltext{1}{Kavli Institute for Astronomy \& Astrophysics and
  Department of Astronomy, Peking University, Yi He Yuan Lu 5, Hai
  Dian District, Beijing 100871, China}

\altaffiltext{2}{Key Laboratory for Optical Astronomy, National
  Astronomical Observatories, Chinese Academy of Sciences, 20A Datun
  Road, Chaoyang District, Beijing 100012, China}

\altaffiltext{3}{Department of Physics and Astronomy, Macquarie
  University, Balaclava Road, Sydney, NSW 2109, Australia}

\altaffiltext{4}{International Space Science Institute--Beijing, 1
  Nanertiao, Zhongguancun, Hai Dian District, Beijing 100190, China}

\altaffiltext{5}{Dipartimento di Fisica e Astronomia `Galileo
  Galilei', Universit\`a di Padova, Vicolo dell'Osservatorio 3,
  Padova, IT-35122, Italy}

\altaffiltext{6}{Research School of Astronomy \& Astrophysics,
  Australian National University, Canberra, ACT 2611, Australia}

\begin{abstract}
Recent discoveries of bimodal main sequences (MSs) associated with
young clusters (with ages $\lesssim 1$ Gyr) in the Magellanic Clouds
have drawn a lot of attention. One of the prevailing formation
scenarios attributes these split MSs to a bimodal distribution in
stellar rotation rates, with most stars belonging to a rapidly
rotating population. In this scenario, only a small fraction of stars
populating a secondary blue sequence are slowly or non-rotating
stars. Here, we focus on the blue MS in the young cluster NGC 1850. We
compare the cumulative number fraction of the observed blue-MS stars
to that of the high-mass-ratio binary systems at different radii. The
cumulative distributions of both populations exhibit a clear
anti-correlation, characterized by a highly significant Pearson
coefficient of $-0.97$. Our observations are consistent with the
possibility that blue-MS stars are low-mass-ratio binaries, and
therefore their dynamical disruption is still ongoing. High-mass-ratio
binaries, on the other hand, are more centrally concentrated.
\end{abstract}


\section{Introduction}\label{sec:intro}

Extended main-sequence turnoff regions (eMSTOs) are common features of
almost all intermediate-age, $\sim$1--2 Gyr-old clusters (IACs) in the
Large and Small Magellanic Clouds (LMC, SMC)
\citep[e.g.,][]{milone2009, girardi2013, li2016, wu2016}. Recently,
such features have been also detected in some young massive clusters
(YMCs; with ages $\leq 300$ Myr) in the LMC
\citep[e.g.,][]{milone2015multiple, milone2016multiple,
  milone2017mnras, li2017}. In addition, some of these YMCs exhibit
clearly split main sequences (MSs), where most cluster stars are
concentrated on or near the cluster population's ridge line in the
color--magnitude diagram (CMD), whereas a smaller fraction of stars
scatter to the ridge line's blue periphery. We will henceforth refer
to these stars as blue-MS stars.

Previous studies have proposed several scenarios to explain the
observed eMSTO regions and the split MSs, including multiple stellar
populations of different ages \citep{milone2009, goudfrooij2011,
  girardi2013, piatti2017}, coeval stellar populations characterized
by different rotation rates \citep{bastian2009, brandt2015,
  D'Antona2015mnras, wu2016}, interacting binaries
\citep{yang2011contributions}, or combinations of these
\citep[e.g.,][]{goudfrooij2015, li2017, milone2017mnras}. Among the
prevailing explanations, differences in stellar rotation rates are
currently deemed most viable. For instance, \citet{D'Antona2015mnras}
showed that the bimodal MS of NGC 1856 can be interpreted as a
superposition of two populations with different rotation rates, where
one population---encompassing two-thirds of all cluster stars---is
characterized by a very high rotation rate of $\omega = 0.9
\omega_{\rm crit}$ (where $\omega_{\rm crit}$ is the break-up angular
velocity) and forms the MS and the upper MSTO region. The second
population contains the remaining one-third of the cluster's stars,
characterized by slowly or non-rotating stars ($\omega = 0$), which
form the blue MS. The underlying idea of the rotational variation
model is that stars with different rotation rates exhibit different
evolutionary behavior for the same stellar mass. For instance, rapid
stellar rotation reduces the stellar surface temperature, causing such
stars to look redder and appear fainter. In addition, convection in
their hydrogen-burning cores would extend their MS lifetime. All of
these effects complicate the resulting morphology of the MS and the
MSTO region.

The stellar rotation model is remarkably successful in explaining the
eMSTO regions and/or bifurcated MSs in most YMCs
\citep[e.g.,][]{brandt2015,
  D'Antona2015mnras}. \citet{D'Antona2015mnras} proposed that the
slowly rotating stellar population may be linked to a cluster's binary
population, since binary interactions slow down stellar rotation
rates. Their proposed stellar rotation scenario therefore provides a
direct link between bifurcated MSs and the binary interaction scenario
\citep{yang2011contributions}.

Binary systems are, on average, more massive than a cluster's stellar
population for the same (primary) mass. Dynamical mass segregation
would therefore cause binaries to gradually sink toward the cluster
center. This picture has been confirmed in most old Galactic globular
clusters \citep[e.g.,][]{milone2012acs}. For YMCs, dynamical
disruption of binary systems should also be taken into account. A
population's `hard' binaries are expected to exhibit a higher degree
of central concentration than single stars, while the `soft' binaries
would be less segregated, because in the central regions of dense
clusters dynamical disruption is more efficient. In this context, we
define soft and hard binaries to satisfy $\vert E \vert / m\sigma^2 <
1$ and $> 1$, respectively, where $E$ is the binary system's binding
energy, and $m\sigma^2 $ is its typical kinetic energy for a combined
mass $m$ and velocity dispersion $\sigma$. \citet{Heggie1975} first
proposed this dichotomy and his predictions have recently been
confirmed \citep{de2013,geller2013,geller2015,li2013}.

Although the radial behavior of binaries in YMCs is complicated, the
different dynamical processes they have experienced relative to single
stars are expected to lead to different radial profiles. A comparison
of the radial distribution of blue-MS stars and binaries may provide
clues about their potential correlation, if any. In this article, we
analyze the radial behavior of the observed blue-MS stars and the
high-mass-ratio binaries in the $\sim 100$ Myr-old, $\sim 4.4 \times
10^4 M_{\odot}$ LMC cluster NGC 1850. The cluster's populations of
blue-MS stars and high-mass-ratio binary systems can be distinguished
easily based on inspection of its CMD \citep[see their
  Fig. 6]{milone2016multiple}.

This paper is organized as follows. In Section 2, we describe the data
analysis processes we adopted. Section 3 presents the main results. We
discuss the physical implications of our results in Section
4. Finally, we provide a summary and preliminary conclusions in
Section 5.

\section{Data Analysis}\label{sec:sample}

We use high-resolution data collected through the Ultraviolet and
Visual Channel of the Wide Field Camera 3 (UVIS/WFC3) on board of the
{\sl Hubble Space Telescope} ({\sl HST}) as part of program GO-14069
(PI: N. Bastian). The exposure times through the F336W, F343N, and
F438W filters are 2555 s, 4075 s, and 1048 s, respectively. Photometry
was done using the DOLPHOT\footnote{DOLPHOT is a stellar photometry
  package for {\sl HST} data developed by Andrew Dolphin. The software
  and the WFC3 module can be found at
  http://purcell.as.arizona.edu/dolphot/.} stellar photometry package
and its WFC3 module. We ran the \textit{wfc3mask},
\textit{splitgroups}, \textit{calcsky}, and \textit{dolphot} tasks in
order, following the preprocessing steps recommended in the {\it
  DOLPHOT/WFC3 User's Guide} to obtain the best photometric
results. The DOLPHOT output files include several parameters to
estimate the quality of our photometry, including
\textit{Signal-to-noise}, \textit{Object sharpness}, \textit{Object
  roundness}, \textit{Crowding}, \textit{Object type}, and
\textit{Photometry quality flag}. In order to restrict our analysis to
the highest-quality photometric data, we use measurements that satisfy
the following five criteria: (1) \textit{signal-to-noise > 5}. (2)
$|\textit{Object sharpness}| <0.2$, which was used to remove
non-stellar objects. The absolute value of an object's sharpness is
small for point-like sources (e.g., stars) that are well fitted by the
PSF model. More positive sharpnesses imply sharper objects (e.g.,
cosmic rays); more negative sharpness corresponds to objects with
broader profiles, such as blended clusters or galaxies. (3)
$\textit{Crowding} <0.5$ to reject stars that are poorly measured
because of contamination by nearby bright objects. (4) $\textit{Object
  type} = 1$, i.e., `good' stars. (5) {\it Photometry quality flag} =
0. The resulting stellar catalog includes a total of 21,660 stars.

Next, we need to define the appropriate cluster and reference
fields. To do so, we first inspected the radial surface brightness
profile of NGC 1850. The spatial distribution and the number-density
contours including all stars are shown in Fig. 1. The red pentagram
indicates the cluster center, which corresponds to the highest
two-dimensional number density. The resulting cluster center
coordinates are $\alpha_{\rm J2000} = 05^{\rm h} 08^{\rm m} 44.34^{\rm
  s}$, $\delta_{\rm J2000} = -68^{\circ} 45' 45.60''$. These
coordinates are in good agreement with the coordinates $\alpha_{\rm
  J2000} = 05^{\rm h} 08^{\rm m} 44.79^{\rm s}$, $\delta_{\rm J2000} =
-68^{\circ} 45' 38.60''$ listed by the Strasbourg Astronomical Data
Center's SIMBAD database (http://simbad.u-strasbg.fr/simbad/). Black
points concentrated around the red pentagram are cluster stars. We
will analyze the behavior of the high-mass-ratio binaries (using their
binary fractions, $f_{\rm bin}$) and blue-MS stars ($f_{\rm bMS}$) as
a function of radius in this region (the extent of the region
  covered here was determined on the basis of the cluster's surface
  brightness profile; see below for details).

At the distance of NGC 1850, $(m-M)_0 =18.45$ mag (determined below
and based on isochrone fitting), $1''$ corresponds to 0.24 pc. As we
will see below, the cluster's core radius, $r_{\rm c}$, is 2.76 pc and
its radial profile disappears into the background noise at a radius of
$\sim$83$''$ or 20 pc. We chose objects located at distances greater
than 30 pc from the cluster center as reference stars (see the red
points in Fig. 1), since those stars should be negligibly contaminated
by cluster stars. They were used to statistically remove background
stars from the adopted cluster region. Additionally, a small young
star cluster (NGC 1850B) located to the southwest of our target
cluster contains several young bright stars. To minimize any
contamination by NGC 1850B, we removed objects located in an area with
a radius of $8''$ centered on NGC 1850B. (The center coordinates of
NGC 1850B were kept consistent with the location of the brightest star
in the raw image, based on visual inspection, because the number
density in this region is too low to determine the subcluster's center
coordinates on the basis of a density contour map. The adopted
coordinates are $\alpha_{\rm J2000} = 05^{\rm h} 08^{\rm m} 39.34^{\rm
  s}$, $\delta_{\rm J2000} = -68^{\circ} 45' 45.80''$.)

\begin{figure}[!htpb]
\begin{center}
\includegraphics[width=0.5\textwidth]{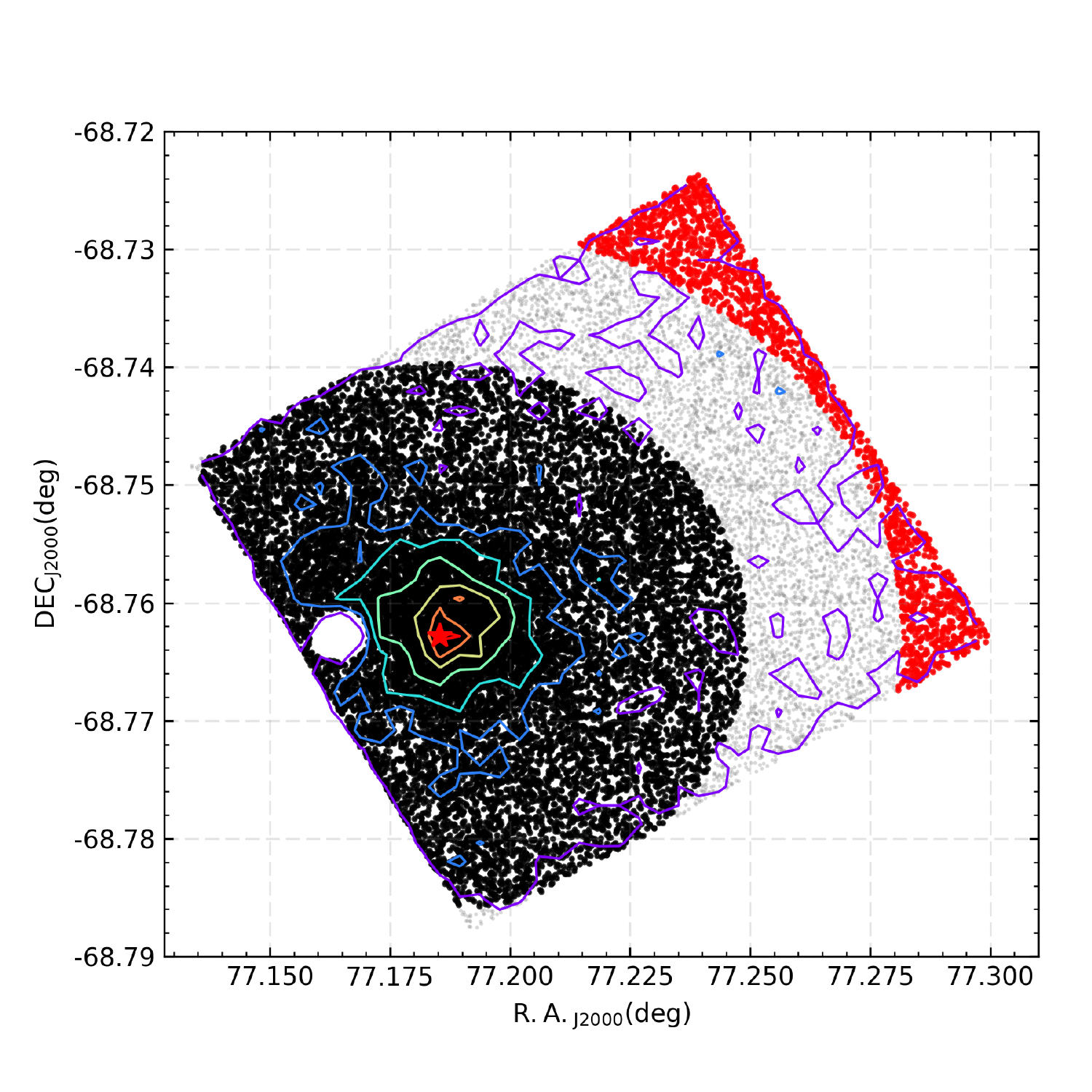}
\caption{Spatial distributions of all stars (grey), stars within the
  adopted cluster region (black), reference field stars (red), and
  number density contours in the NGC 1850 field. The red pentagram
  indicates the cluster center at $\alpha_{\rm J2000} = 05^{\rm h}
  08^{\rm m} 44.34^{\rm s}$, $\delta_{\rm J2000} = -68^{\circ} 45'
  45.60''$. Objects located in the white circle to the left of the red
  pentagram were removed so as to minimize contamination by NGC 1850B
  stars. }
\label{fig:bol1} 
\end{center}
\end{figure}

We used the cluster center to define annular rings at different radii
and calculated the corresponding surface brightness profile. We
followed the method introduced by \citet{mackey2003}, adopting a $4''$
ring width. The resulting surface brightness profile is shown in
Fig. 2. Since NGC 1850 is a YMC in the LMC, its surface brightness
profile follows the canonical \citep[][EFF]{elson1987} profile,
\begin{equation}
\mu(r) = \mu_{0}(1+\frac{r^2}{a^2})^{- \gamma /2},
\end{equation}
where $\mu_0$ is the central surface brightness. The EFF model's core
radius, $a$, and the power-law index $\gamma$ are linked to the core
radius of the standard King model, $r_{\rm c}$, through
\begin{equation}
r_{\rm c} = a(2^{2/\gamma} -1)^{1/2}.
\end{equation}

The parameters pertaining to the best-fitting EFF profile and the
linked core radius of the King profile are also included in the bottom
left-hand corner of Fig. 2. Our fit implies that the NGC 1850 core
radius is $2.76\pm0.12$ pc, which is consistent within one sigma with
the value of $2.69 ^{+0.13}_{-0.17} $ pc derived by \citet[][their
  Table 11]{mclaughlin2005}. Through visual inspection, we determined
that at a radius $R \sim 20$ pc ($\sim$83$''$; indicated by the
vertical red dashed line in Fig. 2) the cluster's surface brightness
profile becomes indistinguishable from the field level. We therefore
adopted this radius as that encompassing the typical cluster region
(shown as the area containing the black points in Fig. 1). Note that
in reality the surface brightness continues to decrease beyond the
cluster region, however, so that the adopted value of 20 pc does not
represent the full extent of the cluster. This is consistent with the
conclusion of \citet{mclaughlin2005}, who derived a tidal radius for
NGC 1850 of $\log (r_{\rm t} \mbox{ pc}^{-1}) = 2.16
^{+0.09}_{-0.07}$.

\begin{figure}
\begin{center}
\includegraphics [width=0.5\textwidth]{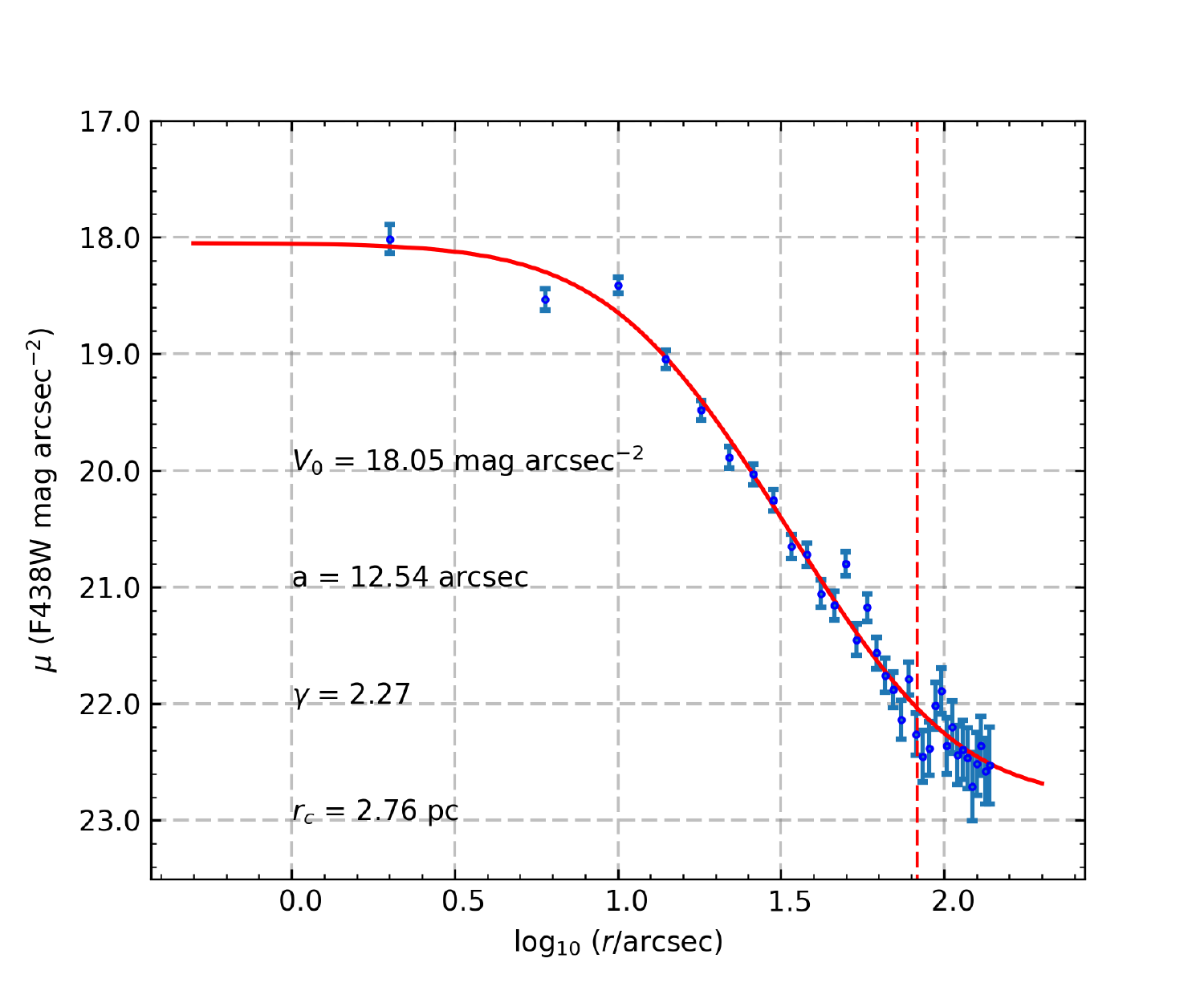}
\caption{Surface brightness profile of NGC 1850. The red solid line is
  the best-fitting EFF profile. Error bars show Poissonian errors
  pertaining to each $4''$-wide annulus. The three parameters
  associated with the best-fitting EFF profile and the core radius
  from the linked King profile are shown in the bottom left-hand
  corner. The vertical red dashed line indicates the radius of the
  adopted cluster region.}
\end{center}
\end{figure}

\section{Results}\label{sec:hst}

The CMDs of all stars in our observed field, the reference field
stars, and the cluster stars are shown in Figs 3a, 3b, and 3c,
respectively. We corrected for differential reddening by application
of the method of \citet{milone2012acs}. The resulting
  differential reddening values across the cluster area ranged from
  $\vartriangle E(B-V) \sim -0.024$ mag to 0.023 mag, adopting the
  relative interstellar extinction coefficients $\frac{A_{\rm
      F336W}}{A_V} = 1.66$ and $\frac{A_{\rm F438W}}{A_V} = 1.33$
  \citep{Girardi2008}. In Fig. 3c, we derived the basic cluster
parameters using isochrone fitting. First, we divided stars with
magnitudes in the range $17 < m_{\rm F438W} <22$ mag into 30 bins. We
then used Gaussian distributions to fit the color histograms in each
bin. We chose the center color values of these Gaussian distributions
as representative of our ridge line, shown as blue squares in
Fig. 3c. We adjusted the parameters of the isochrones with reference
to those blue squares. The best-fitting isochrone from the suite of
PARSEC model isochrones \citep{bressan2012parsec} yielded an age $\log
(t \mbox{ yr}^{-1}) = 7.98^{+0.13}_{-0.15}$, a metallicity $Z=0.008$,
a total extinction $A_V = 0.3 \pm 0.1$ mag, and a distance modulus
$(m-M)_0= 18.45^{+0.06}_{-0.08}$ mag. These parameters are mostly in
good agreement with those derived by \citet{bastian2016young}; these
latter authors found a best-fitting age of 70--140 Myr, $Z=0.008$, and
$(m-M)_0 = 18.35$ mag. Our best-fitting isochrone is shown as the red
line in Fig. 3c. The cluster CMD exhibits a significant split in its
MS in the magnitude range $19.0 <m_{\rm F438W}< 20.5$ mag. A zoomed-in
view of the CMD focused on the split MS (indicated by the blue box in
Fig. 3c) is shown in Fig. 3d. We also performed artificial-star
experiments to make sure that the split MS is a real feature
associated with NGC 1850. We first generated an artificial-star
population containing 1000 data points with the same photometric
properties as each isochrone point. We then reran DOLPHOT by adding
the \textit{Fakestars} option. The simulated CMD is shown in Fig. 3e.

Figure 4 illustrates the process adopted to confirm the
  bimodal MS feature in NGC 1850. Figs 4a and 4b are equivalent to
  Figs 3d and 3e, respectively, showing zoomed-in CMDs of the observed
  and simulated stars in the region where we have detected the split
  MS. In Fig. 4c, we display the cluster's observed stellar
  distribution in the $m_{\rm F438W}$ versus $\vartriangle(m_{\rm
    F336W} - m_{\rm F438W})$ diagram; the latter quantity corresponds
  to the color of the stars minus that of the best-fitting isochrone
  for the corresponding F438W magnitude. Comparing the observed
  $\vartriangle(m_{\rm F336W} - m_{\rm F438W})$ distribution (Fig. 4d)
  with the corresponding simulated distribution (Fig. 4e), it is clear
  that the observed histograms exhibit a significant double peak in
  brighter magnitude bins, which merges into a single peak at fainter
  magnitudes, corresponding to the disappearance of the bimodal
  MS. However, the simulated histograms always shows a single-peaked
  distribution; the larger photometric errors on the faint end broaden
  the width of the peak. We thus conclude that the split MS is
  intrinsic to the cluster and cannot be caused by photometric errors
  alone.

\begin{figure*}[!htpb]
\begin{center}
\includegraphics[width=1.0\textwidth]{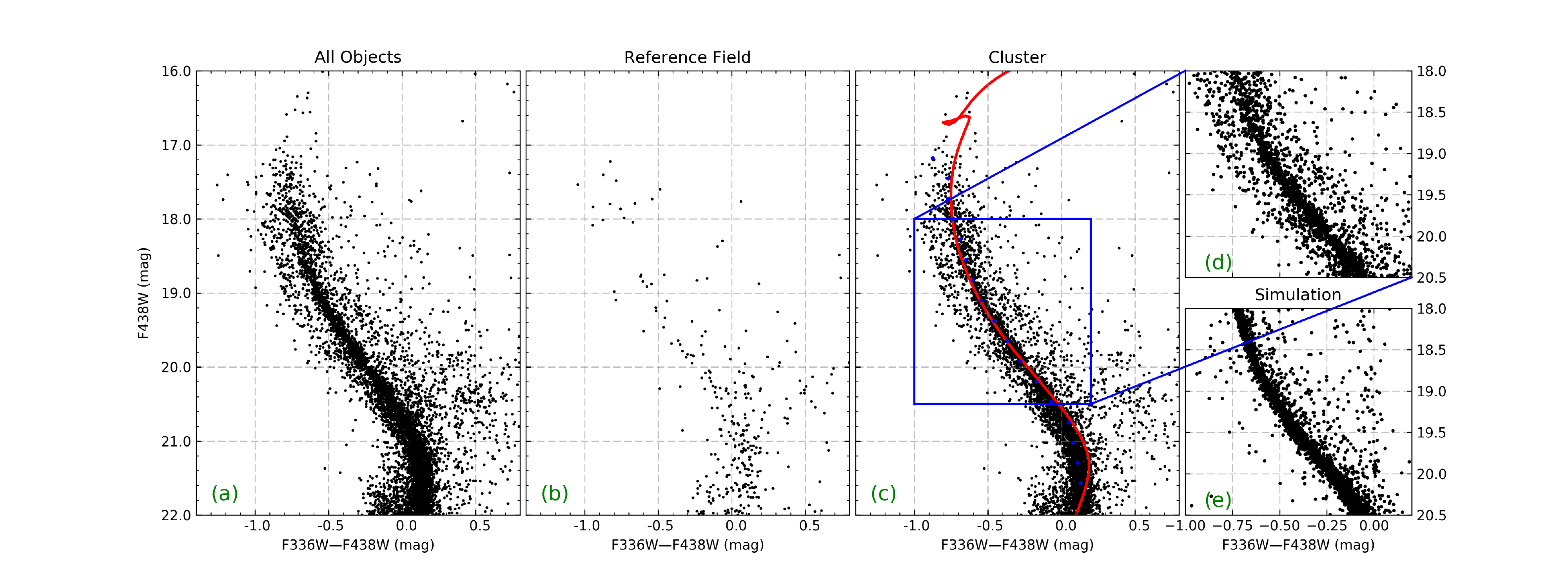}
\caption{CMDs of (a) all stars in our observed field, (b) reference
  field stars (selected in Section 2), and (c) cluster stars. The
  ridge line defined by the cluster stars in the range $17 < m_{\rm
    F438W} < 22$ mag is shown by the blue squares. The red line
  represents the best-fitting isochrone, with $\log(t \mbox{
    yr}^{-1})= 7.98$, $Z=0.008$, $A_V= 0.3$ mag, and $(m-M)_0 = 18.45$
  mag; (d) zoomed-in view focused on the split MS region (indicated by
  the blue box in panel c); (e) CMD of the artificial stars.}
\label{fig:bol1}
\end{center}
\end{figure*}

\begin{figure*}
\begin{center}
\includegraphics[width=1.0\textwidth]{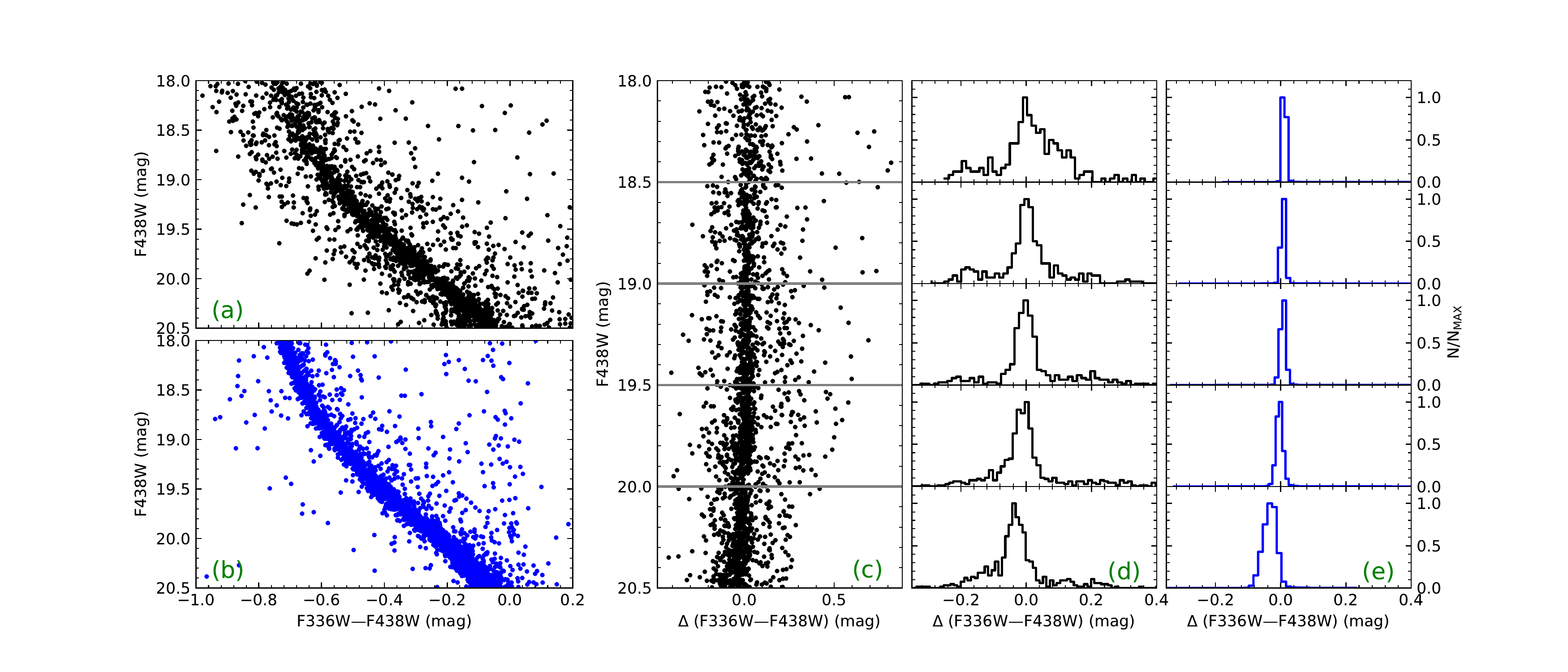}
\caption{Illustration showing that a bimodal MS is an
    intrinsic feature of NGC 1850, which cannot have been caused by
    photometric errors alone. The CMDs of the observed stars (black
    dots) and the simulated stars (blue dots) are shown in panels (a)
    and (b), respectively. Panel (c) shows the $m_{\rm F438W}$ versus
    $\vartriangle(m_{\rm F336W} - m_{\rm F438W})$ CMD based on
    observed data. Panels (d) and (e) show the distribution of the
    $\vartriangle(m_{\rm F336W} - m_{\rm F438W})$ color in five
    magnitude intervals for the observed and simulated stars,
    respectively. }
\end{center}
\end{figure*}

We are interested in the clearly split MS feature located at $19.0
<m_{\rm F438W}<20.5$ mag. Therefore, we applied a method similar to
that adopted by \citet{milone2017mnras} to divide those stars into
blue-MS, red-MS, and high-mass-ratio binary stellar samples. A
zoomed-in view of the CMD and the areas adopted for the blue-MS stars,
red-MS stars, and high-mass-ratio binaries are shown as blue, green,
and pink points in the left-hand panel of Fig. 5. The sample
boundaries were defined as follows. First, we chose the best-fitting
isochrone for the magnitude range $19.0 <m_{\rm F438W}<20.5$ mag. From
the PARSEC output data, i.e., the initial mass and the F336W and F438W
magnitudes, we can then interpolate and derive the magnitude and color
for any binary system for a given mass ratio $q=M_1/M_2$, where $M_1$
and $M_2$ are the masses of the binary system's primary and secondary
components, respectively. Thus, we obtained MS--MS binary sequences
for different mass ratios; for $q=1$, the MS--MS binary sequence
represents a simple upward shift of the single-star isochrone by 0.752
mag \citep[see][their Fig. 3]{elson1998}.

The red and blue boundaries of the red MS are defined by moving the
best-fitting isochrone by an additional $1.5\sigma$ given by the
photometric errors in both the color and magnitude directions. Note
that our data are characterized by small photometric uncertainties,
since we adopted only high-quality data. We also note that the DOLPHOT
output uncertainties are rather too small; if we would use the DOLPHOT
uncertainties directly, the fiducial sequence would not be included in
the red-MS region. Therefore, we adopted the standard deviations of
the magnitudes in both filters from our artificial-star tests
(Fig. 3e) as our photometric uncertainties, which include the
systematic errors. This method is similar to that adopted by
\citet{milone2017mnras}. The latter authors shifted the ridge line of
the cluster's red MS by $2\sigma_{\rm c}$ toward the blue, where
$\sigma_{\rm c}$ is the uncertainty in color.

The bright and faint boundaries of the blue and red MSs were set at
$m_{\rm F438W} = 19.0$ mag and $m_{\rm F438W} = 20.5$ mag,
respectively. The red boundary of the high-mass-ratio binaries was
defined by shifting the sequence of equal-mass binaries to the red by
$1.5\sigma$, where $\sigma$ is the uncertainty in the color. Note that
the pink points are mostly located above the $q=0.55$ MS--MS binary
sequence, hence our reference to `high-mass-ratio' binaries. The top
and bottom boundaries of the pink region were obtained through
interpolation, starting from the different mass ratios corresponding
to the equivalent points on the single-star isochrone at $m_{\rm
  F438W} = 19.0$ mag and $m_{\rm F438W} = 20.5$ mag, respectively. The
blue boundary of the blue MS was drawn arbitrarily to make sure that
we include the majority of objects in that regime. The right-hand
panel of Fig. 5 shows the same boundaries for the reference field
stars. Our final sample of blue-MS stars includes 194 objects; our
catalog also includes 1176 red-MS stars and 210 high-mass-ratio
binaries in the cluster region, as well as three blue-MS stars, 30
red-MS stars, and 10 high-mass-ratio binaries in the reference field.

\begin{figure*}[!htpb]
\begin{center}
\includegraphics[width=1.0\textwidth]{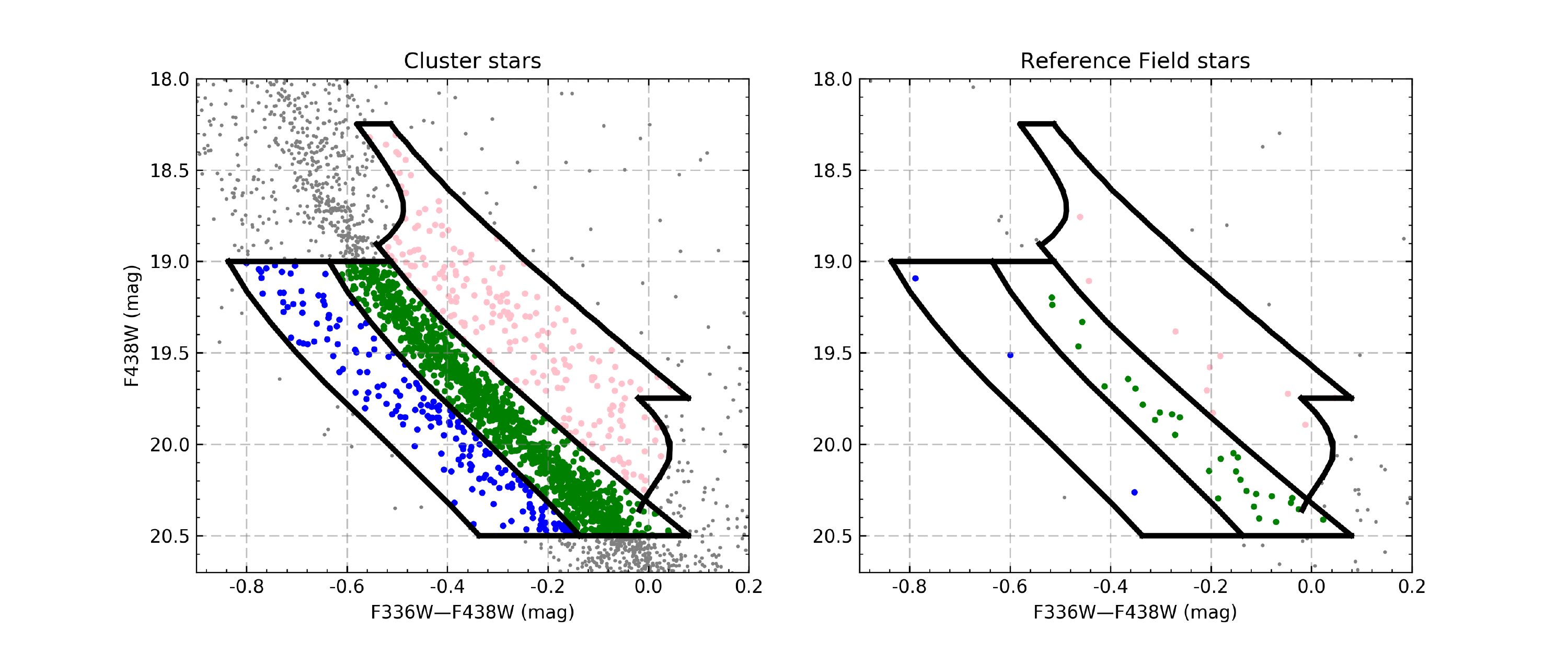}
\caption{Zoomed views of the CMD region of interest exhibiting split
  MSs. The left-hand panel shows the distribution of blue-MS stars
  (blue), red-MS stars (green), and high-mass-ratio binaries (pink) in
  the cluster region. The right-hand panel is the same but for the
  reference field.}
\label{fig:bol1}
\end{center}
\end{figure*}

We calculated the population ratio of blue-MS stars and
high-mass-ratio binaries with respect to the total number of stars in
the three regions for different radii:
\begin{equation}
f_{\rm bMS}(r) = \frac{(N_{\rm bMS}(r) - A(r) n_{\rm bMS})\times
  \frac{P({\rm in})_{\rm bMS}(r)}{P({\rm out})_{\rm bMS}(r)}}{N_{\rm
    bMS}(r) + N_{\rm rMS}(r) + N_{\rm bin}(r)} ;
\end{equation}

\begin{equation}
f_{\rm bin}(r) = \frac{(N_{\rm bin}(r) - A(r) n_{\rm bin})\times
  \frac{P({\rm in})_{\rm bin}(r)}{P({\rm out})_{\rm bin}(r)}}{N_{\rm
    bMS}(r) + N_{\rm rMS}(r) + N_{\rm bin}(r)}.
\end{equation}

Here, $N_{\rm bMS}$, $N_{\rm rMS}$, and $N_{\rm bin}$, are the numbers
of blue-MS stars, red-MS stars, and high-mass-ratio binaries in the
different annular rings, respectively; $n_{\rm bMS}$ and $n_{\rm bin}$
are the corresponding numbers for stars located in the same regions in
the CMD of the reference field. $A(r)$ is the correction factor for
the areal difference between the ring and the reference field. This
value is defined by means of Monte Carlo simulations. (In detail, we
generated one million uniformly distributed data points in the right
ascension--declination plane, counting the numbers of points located
in annular rings and the reference field, i.e, $A(r) = \frac{N_{\rm
    ring}(r)}{N_{\rm reference}}$.)

$\frac{P({\rm in})(r)}{P({\rm out})(r)}$ is a photometric correction
factor. Single stars in the cluster center may appear like binaries
because of the crowded environment in the cluster core and because of
stellar blends. Thus, a blue-MS star can be pushed into the CMD region
of the red-MS stars, while a red-MS star can be pushed into the region
containing the cluster's binaries. It is therefore necessary to
correct for the number difference caused by the process of obtaining
our photometry. To do so, we generated more than 200,000 artificial
stars, inserted them into the raw image, and then measured them in
exactly the same way as the real stars. Specifically, we first
generated 100 artificial stars located in the black-bordered region of
Fig. 5 with uniform color--magnitude and spatial distributions. Next,
we reran DOLPHOT with the added \textit{Fakestars} option, thus
allowing us to assess the differences between the input and output
photometry. We repeated this process more than one thousand times for
each chip to reduce the effects of statistical fluctuations. We
generated 100 artificial stars each time so as to avoid the situation
where artificial stars would significantly increase the crowding of
our images. Finally, we calculated the number ratio of the input and
output artificial stellar samples using the same approach as that
introduced to deal with Fig. 5, thus resulting in the photometric
correction factor.

In this paper, we are only concerned with the number-fraction profiles
of the blue-MS stars and high-mass-ratio binaries. We divided all
colored stars in Fig. 5 into 15 different annular rings, imposing the
condition that these annular rings contain almost the same numbers of
colored stars. The cumulative number-fraction profiles are shown in
Fig. 6. The number fractions of the high-mass-ratio binaries, $f_{\rm
  bin} (\leq R)$, and of the blue-MS stars, $f_{\rm bMS} (\leq R)$,
show a significant negative correlation. The Pearson coefficient
pertaining to the cumulative population is $-0.97$, which satisfies
the condition that ``if the absolute Pearson coefficient exceeds 0.7,
the correlation between two data sequences is significant.'' Figure 6
thus reveals a potential correlation between both stellar samples,
which we will discuss in the next section.

Additionally, radially dependent photometric errors could also
introduce spurious trends. The photometric errors may be large in the
cluster center and small in the outskirts because of crowding in the
cluster core. From the CMD, we infer that more intrinsic red-MS stars
(blue-MS stars, high-mass-ratio binaries) would move to the blue-MS
stars region or high-mass-ratio binaries in the cluster center, while
stars would remain in their original region in the cluster's
periphery. This means that our photometry procedure cannot fully
reproduce the input distribution but introduces a small, non-real
gradient. We conducted artificial-star experiments to check the
importance of this effect on our observed number fraction profiles.
In particular, we generated single fake stars along the ridge lines of
the observed blue- and red-MSs. We also generated 30\% binaries for
each single-star sample with a flat mass-ratio distribution. The
spatial coordinates of all artificial stars follow the same radial
distribution. We obtained PSF photometry for the artificial stars
using DOLPHOT. We used the same approach as employed for the real
stars to reduce the simulated data and derived the radial
distributions of blue-MS stars and binaries. The measured number
fractions of blue-MS stars are significantly higher than the input
values in the inner rings, and they decrease in the outer
rings. However, the measured number fraction of binaries remains
almost the same as the input values in all rings. In other words, the
simulated radial distribution is such that the artificial blue-MS is
more centrally concentrated than the red-MS population. Therefore, we
come to the conclusion that the blue-MS fraction is overestimated,
while radially dependent photometric errors would not change the main
result.
 
\begin{figure}[!htpb]
\begin{center}
\includegraphics[width=0.5\textwidth]{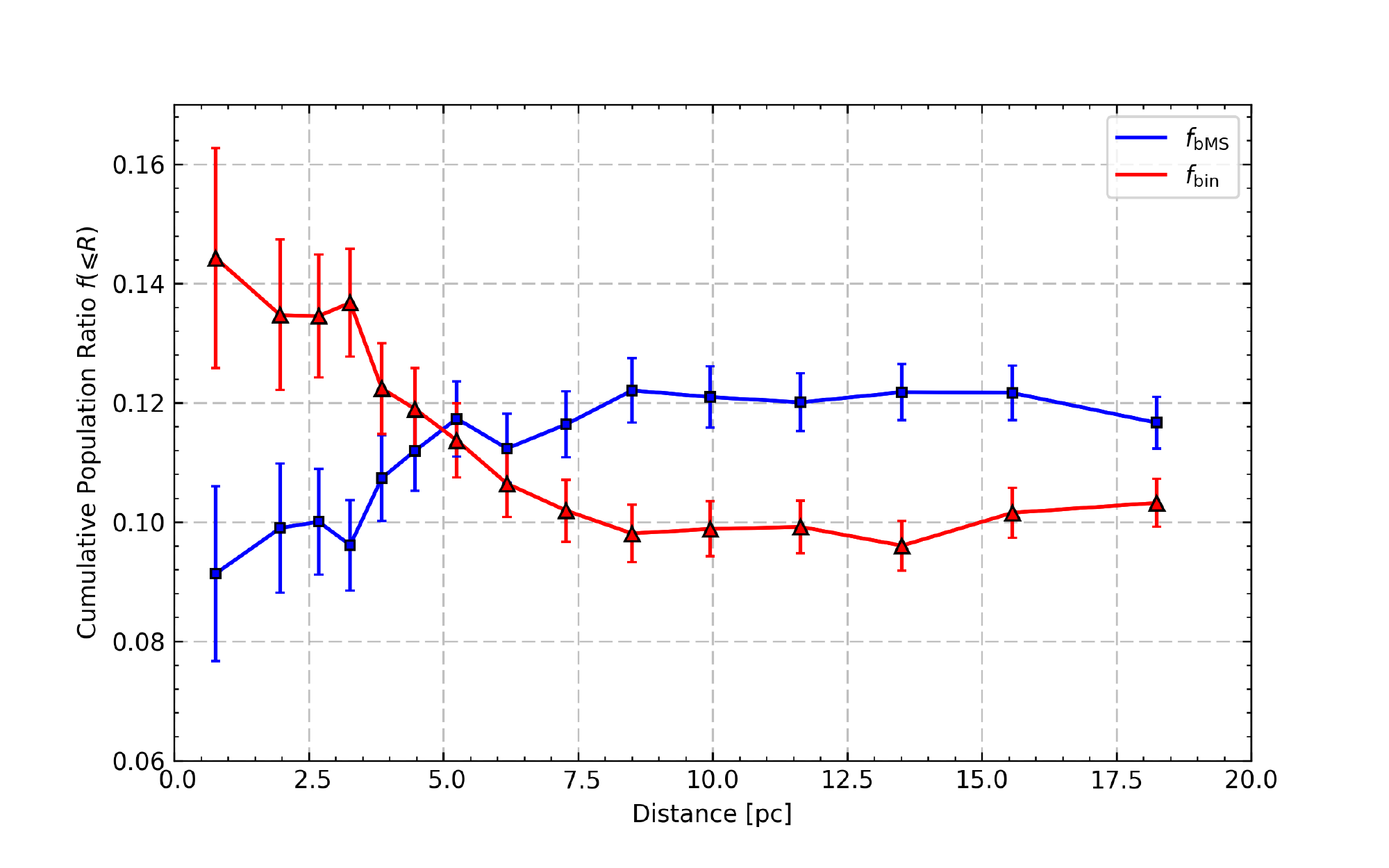}
\caption{Cumulative population of blue MS stars (blue line) and
  high-mass-ratio binaries (red line) with respect to the total
  numbers of sample stars at different radii. The vertical error bars
  represent Poissonian errors. Both profiles are strongly
  anti-correlated for radii within a few parsecs from the cluster
  center.}
\label{fig:bol1}
\end{center}
\end{figure}

\section{Discussion}\label{sec:discuss}

The origin of the blue MS remains a puzzle. \citet{D'Antona2015mnras}
interpreted the CMD of the YMC NGC 1856 as a superposition of two
populations, i.e., one-third of the total number of stars is composed
of slowly/non-rotating stars and two-thirds of rapidly rotating
stars. A key problem arising from this scenario is that rapid rotation
is a common feature of B--A-type stars, so one should then wonder why
there are so many slowly/non-rotating stars. One possibility regarding
the origin of the slowly/non-rotating stars is that most blue-MS stars
might hide a binary component, where the tidal interaction between the
binary components is held responsible for slowing down the rotation
rate \citep{d2017nature}. If this suggestion is correct and also holds
for NGC 1850, there might be a link between the spatial distributions
of the blue-MS stars and the high-mass-ratio binaries. Our results
thus offer some hints as regards this possible correlation.

The origin of the apparent anti-correlation between the radial
number-fraction profiles of the blue-MS stars and the high-mass-ratio
binaries is not clear. \cite{li2013} showed the opposite trend for the
F-type binaries' radial profiles in both NGC 1805 and NGC 1818. They
concluded that this opposite trend could have been caused by the
clusters' different dynamical ages. NGC 1805 is dominated by dynamical
mass segregation while the evolution of NGC 1818 is still dominated by
binary disruption. Their calculations confirmed that NGC 1805 is
dynamically older than NGC 1818. Using $N$-body simulations,
\citet{geller2013} showed that the radial distribution of binary
systems in rich star clusters can evolve from a decreasing trend
toward a cluster's core to an increasing trend, depending on the
dynamical timescale. A cluster's dynamical age is a key parameter of
importance for the shape of a stellar sample's radial profile.

In addition, it seems that most of the binaries populating the blue MS
are low-mass-ratio binaries ($q<0.55$), which we cannot easily
distinguish from single stars, given the prevailing photometric
uncertainties. They are likely still in their dynamical disruption
phase, while the high-mass-ratio binaries are more segregated. To
confirm the dynamical phase governing the blue-MS stars and the
high-mass-ratio binaries in NGC 1850, we calculated the half-mass
relaxation times for both samples \citep{meylan1987},
\begin{equation}
t_{r,{\rm h}} = (8.92\times 10^5)\frac{M_{\rm
    tot}^{1/2}}{m}\frac{R_{\rm h}^{3/2}}{\log(0.4M_{\rm tot}/m)}
\mbox{ yr}.
\end{equation}

The total mass of NGC 1850 was calculated using the relation between
the initial stellar masses and magnitudes in the F438W filter provided
by the best-fitting PARSEC isochrone data table. The total luminous
mass of NGC 1850 based on our F438W observations is $\log M_{\rm
  tot}/M_{\odot} = 4.360\pm 0.001$ and the half-mass radius $R_{\rm h}
= 9.239^{+0.009}_{-0.017}$ pc. We also estimated the mass of NGC 1850
based on the canonical initial mass function \citep{kroupa2001}, $\log
M_{\rm tot}/M_{\odot} = 4.64^{+0.09}_{-0.07}$.
\citet{correnti2017dissecting} derived $\log M_{\rm tot}/M_{\odot} =
4.86\pm0.10$ and $\log M_{\rm tot}/M_{\odot} = 4.62\pm 0.10$ based on
\citet{salpeter1955} and \citet{chabrier2003} initial mass functions,
respectively. The luminous mass estimate we obtained here is the
smallest value. The typical masses of our stellar samples are also
required to calculate the half-mass relaxation time. Since binaries
are completely unresolved in the LMC, binary systems will look like
single point-like sources. With $m_1$, $m_2$, $F_1$, and $F_2$ defined
as the magnitudes and fluxes of primary and secondary stars in a
binary system, the combined magnitude of the binary system is
\begin{equation}
m_{\rm bin} = m_1 - 2.5 \log(1+\frac{F_2}{F_1}).
\end{equation}

Since the observed fluxes are related to the underlying
  stellar masses, the combined magnitudes of MS--MS binary systems
  depend on the relevant mass ratios, $q=M_2/M_1$. We can calculate
the MS--MS binary sequences for different mass ratios and the curve
defining the change in luminosity for a given initial mass
\citep{elson1998}. MS--MS binary sequences and luminosity change
curves compose a grid covering the CMD. In this grid, one object
corresponds to one set of coordinates, $(q,M_{\rm ini})$. The mass of
photometric MS--MS binary systems can be obtained by optimizing the
combination of $q$ and $M_{\rm ini}$. We thus calculated the average
mass of our high-mass-ratio binaries, i.e. $4.52 M_{\odot}$. The
average mass of blue-MS stars is $2.52 M_{\odot}$, which was
calculated through interpolation (i.e, we can derive the mass of a
blue-MS star with any magnitude through the relationship between the
initial mass and the F438W magnitudes of the best-fitting isochrone
data).
 
Our aim of estimating the half-mass relaxation time here is to compare
the dynamical stages of our stellar samples, not to determine precise
values. Our rough estimates of the half-mass relaxation times of the
high-mass-ratio binaries and the cluster's blue-MS stars are 253.7 Myr
and 422.6 Myr, respectively, while the respective core relaxation
times are 38.3 Myr and 63.8 Myr. The dynamical timescale is always
longer for the blue-MS stars than for the high-mass-ratio
binaries. This confirms the assumption that the blue-MS stars are
dynamically younger than the high-mass-ratio binaries. This then
supports the suggestion that the blue-MS stars are low-mass-ratio
binaries, which have lower binding energies than high-mass-ratio
binary systems, because the binding energy of a binary system is
directly proportional to the masses of its stellar components ($E
\varpropto q M^2$), on average. Generally speaking, after suffering
numerous encounters in a stellar system, soft binaries become softer,
while hard binaries become harder \citep{Heggie1975}. That is indeed
the reason why the high-mass-ratio binaries in NGC 1850 are more
segregated than their low-mass-ratio counterparts. However, ideally we
prefer a shorter half-mass relaxation timescale for high-mass-ratio
binaries than the cluster age, while we would expect the equivalent
timescale for blue-MS stars to be longer than the cluster age. Of
course, this discussion assumes that the cluster is located in an
isolated environment. We note, however, that NGC 1850 is located at
the northern end of the LMC's bar structure. In addition, the young
cluster NGC 1850B is located to the west of NGC 1850, within about
$30''$. Both of those structures may contribute to accelerating the
cluster's dynamical evolution, so the half-mass relaxation times
estimated above are upper limits.
 
We tentatively suggest that dynamical mass exchange of binaries may
link the low- and high-mass-ratio binaries, thus producing a radial
anti-correlation between them. There are two channels by which a
binary system can be hardened through interactions with single
stars. First, extraction of internal energy from the binary by a
single star would make a binary system more strongly bound. This
process is important when the average mass of the surrounding stars
and the local stellar number density are not very high, e.g., in the
outer regions of star clusters. Second, a low-mass-ratio binary system
can transfer internal energy to its surroundings through mass
exchange. The lower-mass component in the binary system is thereby
replaced by a higher-mass field star. As a result, it becomes a
high-mass-ratio binary system. This process is favored when the masses
of the surrounding stars are higher than the mass of one component of
a binary system, e.g., in the mass-segregated core region of a star
cluster.

The second channel for binary hardening is actually a combination of
the destruction of low-mass-ratio binaries and the production of
high-mass-ratio binaries, thus causing mutual exclusion. Because all
binaries studied here have mass ratios $q \geq 0.55$, if the blue-MS
stars are indeed low-mass-ratio binaries and if they are still
affected by mass exchange, their radial number-fraction profile may be
anti-correlated with the high-mass-ratio binaries. In addition,
because mass exchange favors dense cluster core regions, the
anti-correlation between the number fraction of low- and
high-mass-ratio binaries should be more significant in a cluster's
central region, which is indeed observed in NGC 1850 (Fig. 6).

Our proposed scenario agrees to a large extent with the
  stellar rotation models. However, here we suggest that the blue-MS
  may be populated by low-mass-ratio binaries, whereas
  \citet{d2017nature} interpreted blue-MS stars as slowly rotating
  stars following a period of `braking.' Indeed, it is well-known that
  rapid rotation is a common feature among stars, and tidal torques in
  binary systems can slow down the rotation rates significantly. There
  is no direct observational evidence showing that low-mass-ratio
  binaries are more likely to be affected by tidal interactions than
  high-mass-ratio binaries. For the same gravitational environment and
  conditions, low-mass-ratio binaries preferentially have smaller
  separations. Thus, they can survive just as well as high-mass-ratio
  binaries, since the binding energy $E \varpropto
  \frac{qM^2}{r}$. The smaller the binary system's separation is, the
  stronger the effects of tidal synchronization will be. Therefore,
  the rotation rates of the primary stars of surviving low-mass-ratio
  binaries are more easily affected by tidal synchronization. As a
  result, we would observe bluer primary stars on the whole, because
  of the decrease in rotation rates. In addition, if tidal
  synchronization slows down the primary stars in unresolved systems,
  high-mass companions would make stars look redder and brighter, and
  high-mass-ratio binaries would not be found on the blue-MS. Note
  that, at the present time, this is merely a speculative conclusion.

However, \citet{milone2017mnras} studied the same behavior for another
YMC, NGC 1866. They did not detect any evidence of such an
anti-correlation between the radial number-fraction profiles of the
blue-MS stars and the cluster's high-mass-ratio binaries. On the
contrary, their result seems to support that the number-fraction
profiles of the blue-MS stars and the high-mass-ratio binaries are
positively correlated (their Fig. 9). The half-light relaxation times
of NGC 1866 and NGC 1850 are $\log t_{r,{\rm h}} =
9.55_{-0.08}^{+0.06}$ yr and $\log t_{r,{\rm h}} =
9.58_{-0.10}^{+0.05}$ yr, respectively \citep{mclaughlin2005}. The
best-fitting isochrone age for NGC 1866 varies from 140 Myr to 220
Myr. Comparing the cluster ages and half-mass relaxation times of
these two YMCs, we come to the conclusion that NGC 1866 is dynamically
much older than NGC 1850, which again supports our proposed
scenario. In summary, NGC 1866 is dynamically more evolved than NGC
1850, and the pool of low-mass-ratio systems may already have evolved
in the core of NGC 1866. In turn, this has resulted in the observed
positive correlation between this cluster's blue-MS and
high-mass-ratio radial profiles.

Because measuring the mass ratio of an individual binary system
located in a dense LMC cluster is not possible at the present time,
our proposed scenario is currently only a theoretical
possibility. Similar studies for other analogous YMCs will help us to
better understand the potential physics governing these blue-MS stars.

\section{Summary}\label{sec:sum}

In this paper, we have studied the radial number-fraction profiles of
the blue-MS stars and the high-mass-ratio binaries in the YMC NGC
1850. We aimed to examine if the observed blue-MS stars are
binary-related objects, as suggested by \citet{D'Antona2015mnras,
  d2017nature}. Our analysis reveals that (1) the blue-MS stars show
an inverse radial segregation, while the high-mass-ratio binaries are
more segregated than the cluster's bulk stellar population; and (2)
the radial number-fraction profiles of the blue-MS stars and the
high-mass-ratio binaries are strongly anti-correlated in the cluster
core.

We suggest that most blue-MS stars may be low-mass-ratio
binaries. They might still be experiencing dynamical disruption. Mass
exchange between these low-mass-ratio binaries and the surrounding
population of massive stars is likely taking place at the present
time, which causes an increase in their mass ratio. This process is
responsible for turning low-mass-ratio binaries into high-mass-ratio
systems, thus producing an inverse correlation between both types of
binary system. However, because of the absence of definitive evidence,
whether or not this scenario is correct remains an open question.

\acknowledgments This work was supported by the National Key Research
and Development Program of China through grant 2017YFA0402702. We also
acknowledge research support from the National Natural Science
Foundation of China (grants U1631102, 11373010, and 11633005).  We thank anonymous referee for several suggestions that have improved the quality of this manuscript. C.L. is supported by the Macquarie Research Fellowship Scheme. A.P.M
acknowledges support from the Australian Research Council through a
Discovery Early Career Researcher Award, DE150101816. A.P.M acknowledges 
support by the European Research Council through the ERC-StG 2016 project 
716082 `GALFOR'.

\bibliography{ngc1850.bib}

%\clearpage

\end{document}